\documentclass{PoS}

\usepackage{amsfonts}
\usepackage{amsmath}
\usepackage{subfigure}

\title{One loop matching factors for improved 
  staggered four-fermion operators with improved glue}

\ShortTitle{Matching factors for staggered four-fermion operators
}

\author{\speaker{Jongjeong Kim}, \ \  Weonjong Lee\\
  Lattice Gauge Theory Research Center, FPRD, and CTP \\
  Department of Physics and Astronomy, 
  Seoul National University, Seoul, 151-747, South Korea \\
  E-mail: \email{rvanguard@phya.snu.ac.kr}, \ \  
          \email{wlee@snu.ac.kr}}

\author{Stephen R. Sharpe\\
  Physics Department, University of Washington, Seattle, WA 98195-1560 \\
  E-mail: \email{sharpe@phys.washington.edu}}

\abstract{We present results for matching factors for staggered
  four-fermion operators constructed using HYP-smeared fat links
  both in the action and the operators.
 We use perturbation theory to calculate the matching factors and work
  to one-loop order. The new feaure of this work is the use of
  the Symanzik-improved gauge action, as opposed to the Wilson gauge action.
 Our results are needed for our ongoing calculation of weak matrix elements
 using HYP-smeared staggered valence quarks and operators on MILC lattices.
 We give explicit results for matching factors of
the operator needed to calculate $B_K$.
 We compare the impact of the improvement of the
 gauge action on one-loop coefficients with that of mean-field improvement
 of the operators.}

\FullConference{The XXVIII International Symposium on Lattice Field 
  Theory, Lattice2010\\
  June 14-19, 2010\\
  Villasimius, Italy}

\begin{document}

\section{Introduction}

Matrix elements of electroweak four-fermion operators between hadronic
states play a central role in constraining the parameters of the CKM
matrix.  The Wilson coefficients of the operators are typically
calculated in a continuum renormalization scheme (usually the NDR
[Naive Dimensional Regularization] scheme with $\overline{\rm MS}$
subtraction), so one needs to convert the matrix elements calculated
on the lattice into this continuum scheme. The multiplicative factors
that are needed (which are, in general, matrices) are called matching
factors or ``Z-factors''. We consider here the matching factors for
four-fermion operators constructed using HYP-smeared staggered
fermions \cite{Hasenfratz:2001hp}, and calculate them at one-loop
order in perturbation theory.  These are needed for our ongoing
calculation of $B_K$ using HYP-smeared staggered
fermions~\cite{Bae:2010ki} on the MILC ``asqtad''
lattices~\cite{milc-01-1}.

The new feature compared to the previous calculation~\cite{Lee:2003sk}
of four-fermion matching factors is the use of the Symanzik-improved gluon
action instead of the Wilson plaquette action. This extension is
necessary because the MILC lattices are generated using the former action.

\section{Action and Feynman Rules}

For valence quarks we use the HYP-smeared action, which is simply
the unimproved staggered action 
\begin{equation} 
  S_f = 
  \frac{1}{2} \sum_{n, \mu} \eta_{\mu}(n) \bar{\chi}(n) 
  \Big( 
  V_{\mu}(n) \chi(n + \hat{\mu}) - 
  V^{\dagger}_{\mu}(n - \hat{\mu} ) \chi(n - \hat{\mu}) \Big)
  + m \sum_{n} \bar{\chi}(n)\chi(n) 
  \,, 
\end{equation}
(where $\eta_\mu(n) = (-1)^{n_1 + \cdots + n_{\mu-1}}$,
and we have set the lattice spacing $a$ to unity)
except that the links, 
$V_\mu$, are HYP-smeared fat links.
By contrast, in the unimproved action the links are replaced by
the original, ``thin'', links, $U_\mu$.
The relation between these two types of links is given in 
Ref.~\cite{Hasenfratz:2001hp}.
What we need is the relation in perturbation theory and in momentum space:
\begin{eqnarray}
  V_\mu(n) &=& \exp(ig B_\mu(n)) \\
  B_\mu(n) &=& \int^{\pi}_{-\pi} \frac{d^4k}{(2\pi)^4} \sum_\nu
  h_{\mu\nu}(k) A_\nu(k) e^{ik\cdot(n + \hat{\mu}/2)} + \mathcal{O}(A^2) \,.
  \label{eq:V_mu_series}
\end{eqnarray}
Here $A_\nu$ are the momentum space gluon fields obtained from
the unimproved links,
and $h_{\mu\nu}(k)$ is a smearing kernel containing all relevant information
about the HYP smearing. For the HYP links whose coefficients are
chosen to remove $\mathcal{O}(a^2)$ taste symmetry breaking at tree
level ($\alpha_1=0.875$, $\alpha_2=4/7$ and $\alpha_3=0.25$
in the notation of Ref.~\cite{Hasenfratz:2001hp}),
$h_{\mu\nu}(k)$ is~\cite{Lee:2002ui,Kim:2009tk,Kim:2010fj}
\begin{equation}
  h_{\mu\nu}(k) = \delta_{\mu\nu} D_\mu(k) + (1 -
  \delta_{\mu\nu}) \bar{s}_\mu \bar{s}_\nu \tilde{G}_{\nu,\mu}(k) \,,  
\end{equation}
where
\begin{align}
  D_\mu(k) &= 1 -
  \sum_{\nu\ne\mu} {\bar s}_\nu^2 + \sum_{\nu < \rho \atop
    \nu,\rho\ne\mu}{\bar s}_\nu^2 {\bar s}_\rho^2 - {\bar s}_\nu^2
  {\bar s}_\rho^2 {\bar s}_\sigma^2 \,, \\
  \tilde{G}_{\nu,\mu}(k) & =  1 - \frac{({\bar s}_\rho^2 
      + {\bar s}_\sigma^2)}{2} 
    + \frac{{\bar s}_\rho^2 {\bar s}_\sigma^2}{3} \,. 
\end{align}
Here, $\mu \ne \nu \ne \rho \ne \sigma$ and $\bar{s}_\mu =
\sin(k_\mu/2)$. We recover the unimproved action
(in which the links are thin) by setting
$h_{\mu\nu}(k) = \delta_{\mu\nu}$.

One would expect the $\mathcal{O}(A^2)$ term
in Eq.~(\ref{eq:V_mu_series}) to be needed at one-loop order,
since it can give rise to tadpole diagrams.
It turns out, however, that its contribution
vanishes due to the $SU(3)$ projection used in HYP-smearing,
as discussed in Refs.~\cite{Patel:1992vu,Kim:2009tk,Lee:2002fj}.

We use the tree level Symanzik-improved gluon
action~\cite{Weisz:1982zw,Luscher:1984xn}, which can be written as
\begin{equation}
  S_g = \frac{2}{g_0^2} 
  \bigg[ \frac{5}{3} \sum_{\rm pl} {\rm ReTr} (1 - U_{\rm pl}) 
  - \frac{1}{12} \sum_{\rm rt} {\rm ReTr} (1 - U_{\rm rt})  
  \bigg] \,,
  \label{eq:sg} 
\end{equation} 
where ``pl'' and ``rt'' 
represent plaquette and rectangle, respectively.
In fact, the MILC collaboration use a (partially) one-loop improved
Symanzik gluon action to generate their configurations~\cite{Alford:1995hw},
and not the tree-level form.
However, the one-loop corrections in the gluon action do not impact
the matching factors of fermionic operators until two-loop level,
and thus it is sufficient to use the tree-level gluon action.
Similarly, the effect of dynamical quarks does not enter until
two-loop level, so the fact that the valence and sea quarks have different
actions does not impact our one-loop calculation.

Feynman rules for the staggered action and four-fermion operators can be found in
literature, and we do not repeat them here. Quark propagators and
vertices can be found in
Refs.~\cite{DS88,Ishizuka:1993fs,Patel:1992vu}. The generalization to
various improved staggered fermions such as asqtad and HYP-smeared staggered
fermions can be found in Ref.~\cite{Lee:2002ui,Lee:2003sk}.  The
improved gluon propagator can be found in
Refs.~\cite{Weisz:1982zw,Kim:2009tk,Kim:2010fj}.

\section{Four-fermion Operators}

We use the hypercube construction of Ref.~\cite{KlubergStern:1983dg}
for our staggered four-fermion operators. 
These come in two kinds, which differ in the contractions of
their color indices.
First we have ``one color-trace'' operators, labeled with a subscript
$I$, whose form is
\begin{align}
  [S \times F][S' \times F']_I (y) =
  \frac{1}{4^4}\sum_{A,B,A',B'}
  & [\bar{\chi}_a^{(1)}(2y+A)
  \overline{(\gamma_S \otimes \xi_F)}_{AB}
  \chi_b^{(2)}(2y+B)] \nonumber\\
  \times 
  & [\bar{\chi}_{a'}^{(3)}(2y+A')
  \overline{(\gamma_{S'} \otimes \xi_{F'})}_{A'B'}
  \chi_{b'}^{(4)}(2y+B')] \nonumber\\
  \times 
  & \mathcal{V}^{ab'}(2y+A,2y+B')
  \mathcal{V}^{a'b}(2y+A',2y+B) \,.
\end{align}
Here $y \in \mathbf{Z}^4$ is the coordinate of hypercubes,
hypercube vectors $S$ and
$S'$ denote the spins of the component bilinears,
while $F$ and $F'$ denote the tastes. 
Indices $a$, $b$, $a'$, and $b'$ denote colors,
while superscripts $(i)$ for $i=1,2,3,4$ label different flavors
(not tastes). Different flavors are chosen to
forbid penguin diagrams. Two fat Wilson lines
$\mathcal{V}^{ab'}(2y+A,2y+B')$ and $\mathcal{V}^{a'b}(2y+A',2y+B)$
ensure the gauge invariance of the four-fermion operators. A fat
Wilson line $\mathcal{V}^{ab'}(2y+A,2y+B')$, for example, is
constructed by averaging over all the shortest paths connecting $2y+A$
and $2y+B'$, with each path formed by products of HYP-smeared links
$V_\mu$. When we use the unimproved staggered action the Wilson lines
are composed of unsmeared links, $U_\mu$.

The second kind of operator is called ``two color-trace'',
and takes the form
\begin{align}
  [S \times F][S' \times F']_{II} (y) =
  \frac{1}{4^4}\sum_{A,B,A',B'}
  & [\bar{\chi}_a^{(1)}(2y+A)
  \overline{(\gamma_S \otimes \xi_F)}_{AB}
  \chi_b^{(2)}(2y+B)] \nonumber\\
  \times 
  & [\bar{\chi}_{a'}^{(3)}(2y+A')
  \overline{(\gamma_{S'} \otimes \xi_{F'})}_{A'B'}
  \chi_{b'}^{(4)}(2y+B')] \nonumber\\
  \times 
  & \mathcal{V}^{ab}(2y+A,2y+B)
  \mathcal{V}^{a'b'}(2y+A',2y+B') \,,
\end{align}
where the subscript $II$ denotes two color-traces. 
These operators differs from those with  one color-trace by the choice
of fat Wilson lines---here they connect within each bilinear,
whereas for the one color-trace operators they connect between bilinears.

Following Refs.~\cite{Patel:1992vu,Ishizuka:1993fs,Lee:2001hc}, we also
consider mean-field improved operators. For the unimproved action
and operators this is usually referred to 
as tadpole improvement~\cite{Lepage:1992xa}.
For the HYP-smeared action and operators, mean-field improvement is
achieved by rescaling
the staggered fields and the link matrices:
\begin{equation}
  \chi \to \psi = \sqrt{v_0} \chi \,,
  \qquad
  \bar{\chi} \to \bar{\psi} = \sqrt{v_0} \bar{\chi} \,,  
  \qquad
  V_\mu \to \tilde{V}_\mu = V_\mu / v_0 \,,
\end{equation}
where $v_0$ is the mean-field improvement factor defined by
\begin{equation}
  v_0 \equiv \bigg[ \frac13 \text{ReTr} \langle V_{\rm pl}
  \rangle\bigg]^{1/4} \,,
\end{equation}
with $V_{\rm pl}$ the plaquette composed of HYP-smeared links.

\section{Renormalization of Four-fermion operators}

Analytic formulae for one-loop perturbative corrections to
the four-fermion operators for HYP-smeared staggered fermions
with the Wilson plaquette action are given in Ref.~\cite{Lee:2003sk}.
It turns out that the generalization to using the improved gluon action
rather than the Wilson plaquette action can be achieved simply by
replacing the ``composite'' gluon propagator;
\begin{equation} 
  (1/\hat{k}^2)
  \sum_\lambda h_{\mu\lambda}h_{\nu\lambda} \to
  \sum_{\alpha\beta} h_{\mu\alpha}h_{\nu\beta} {\cal D}_{\alpha\beta}\,,
\end{equation}
where ${\cal D}_{\alpha\beta}$ is the improved gluon propagator, while
the Wilson gluon propagator is $(\delta_{\alpha\beta}/\hat{k}^2)$ with
$\hat{k}^2 = 4\sum_\alpha \sin^2 (k_\alpha/2)$. This simplification
holds both for bilinear operators~\cite{Kim:2009tk} and four-fermion
operators.

The Feynman diagrams which contribute to one-loop matching factors for
four-fermion operators can be found in Ref.~\cite{Lee:2003sk}.
We have undertaken two independent calculations, using significantly
different methods, to cross-check our results.
Another check is obtained by using Fierz transformations,
which interchange one and two color-trace operators.
The form of these transformations relevant to our set-up is
\begin{equation}
  \overline{(\gamma_S \otimes \xi_F)}_{AB}
  \overline{(\gamma_{S'} \otimes \xi_{F'})}_{A'B'}
  = \frac{1}{16}
  \sum_{MN}  
  \overline{(\gamma_S \gamma_M^\dagger
   \otimes \xi_N^\dagger \xi_{F'})}_{AB'}
  \overline{(\gamma_{S'} \gamma_M
   \otimes \xi_N \xi_F)}_{A'B} \,,
\end{equation}
where $M$, $N$, $A$, $A'$, $B$, and $B'$ are
hypercube vectors. 

The result of the calculation can be expressed as
\begin{equation}
  \vec{O}_i^\text{Cont, (1)}(\mu)
  = \sum_j Z_{ij}(\mu, a)
  \vec{O}_j^\text{Lat}(1/a) \,,
\end{equation}
where the superscripts ``Cont'' and ``Lat'' represent the continuum
and lattice respectively. The sub-indices $i$ and $j$ run over all
combinations of spins and tastes of the four-fermion operators
that are allowed by the lattice symmetries. The
vectors on the four-fermion operators indicate that one and
two color-trace operators, for given spins and tastes, are collected
into a two-dimensional vector. We call the vector space in
which these vectors live the ``color-trace'' space. The renormalization
scale in the continuum scheme is $\mu$, and $a$ (now reintroduced)
is the lattice spacing. Finally, the superscript ``(1)'' indicates
that the continuum operator has been matched at one-loop order.

The matching factor $Z_{ij}$ takes the form
\begin{equation}
  Z_{ij} = \delta_{ij} +
  \frac{g^2}{(4\pi)^2}
  \bigg[
  - \hat{\gamma}_{ij} \log (\mu a) + \hat{c}_{ij}
  \bigg] \,.
\end{equation}
where $\hat{\gamma}_{ij}$ and $\hat{c}_{ij}$ are, respectively,
the anomalous dimension matrix and the finite coefficients.
Apart from their explicit indices,
both are also matrices in the color-trace space, as denoted by
the ``hats''. The finite coefficients
are given by the difference of finite terms in the
continuum and lattice one-loop calculations,
\begin{equation}
  \hat{c}_{ij} 
  = \hat{c}_{ij}^\text{Cont}
  - \hat{c}_{ij}^\text{Lat} \,.
\end{equation}

\section{Matching Factors for $B_K$}

We have obtained the matching factors for all four-fermion operators
of the forms given above, i.e. for all choices of $S$, $S'$, $F$ and
$F'$. Detailed results will be presented elsewhere~\cite{inprep}.
We consider here only the matching factors needed for $B_K$. The continuum
four-fermion operator relevant to $B_K$ is
\begin{equation}
  \mathcal{O}^{Cont}_{B_K} =
  [\bar{s}^a \gamma_\mu  (1-\gamma_5) d^a] 
  [\bar{s}^b \gamma_\mu  (1-\gamma_5) d^b] \,.
\end{equation}
The corresponding operator on the lattice, in the ``two spin-trace''
formulation~\cite{Sharpe:1986xu,kilcup}, can be written at tree-level as
\begin{equation}
  \mathcal{O}_{B_K}^\text{Lat}
  \equiv \sum_{i=1}^4 \mathcal{O}^\text{Lat}_i
  = \mathcal{V}_I
  + \mathcal{V}_{II}
  + \mathcal{A}_{I}
  + \mathcal{A}_{II} \,,
\end{equation}
where
\begin{align}
  \mathcal{O}^\text{Lat}_1 \equiv 
  \mathcal{V}_I = [V_\mu \times P][V_\mu \times P]_I \,,
  & \qquad
  \mathcal{O}^\text{Lat}_2 \equiv 
  \mathcal{V}_{II} = [V_\mu \times P][V_\mu \times P]_{II} \,,
  \\
  \mathcal{O}^\text{Lat}_3 \equiv 
  \mathcal{A}_I = [A_\mu \times P][A_\mu \times P]_I \,,
  & \qquad
  \mathcal{O}^\text{Lat}_4 \equiv 
  \mathcal{A}_{II} = [A_\mu \times P][A_\mu \times P]_{II} \,.
\end{align}
Here, $V_\mu$, $A_\mu$, and $P$ stand for the vector, axial-vector,
and pseudo-scalar respectively. The taste of all the operators
are taken to be the pseudo-scalar $P$, i.e. $F = \xi_5$. 

At one-loop order, many lattice operators contribute to the matching
formula. In practice, however, in our numerical calculation of
$B_K$ we keep only those operators in which the bilinears have
taste $\xi_5$.\footnote{%
The rationale for this is that we use external kaons with taste $\xi_5$.
As shown in Ref.~\cite{ruth_steve}, however, leaving out the operators with
other tastes leads to an error of ${\cal O}(\alpha_s m_K^2/\Lambda_\chi^2)$,
which is of next-to-leading order in staggered chiral perturbation theory.
This error must be accounted for when fitting.}
It turns out that only the four operators listed
above have this taste. Thus what is relevant for our numerical calculation
is the truncated one-loop matching formula
\begin{equation}
  \mathcal{O}_{B_K}^\text{Cont, (1), Trunc}
  = \sum_{i=1}^4 c_i
  \mathcal{O}^\text{Lat, (1)}_i \,,
  \qquad 
  c_i \equiv \Bigg[ 
  1 + \frac{g^2}{(4\pi)^2}
  \bigg( - 4 \log (\mu a)
  + [-\frac{11}{3} - c^\text{Lat}_i]
  \bigg)
  \Bigg] \,.
  \label{eq:mf_BK}
\end{equation}
We use here the NDR scheme in the continuum, for which the finite
coefficient is $-\frac{11}{3}$ for all four operators.

Results for $c^\text{Lat}_i$ are given in
table~\ref{tab:ff_cLat} for various choices of lattice operator.
We see that, in most cases, the coefficients are slightly reduced
by improving the gauge action, although the reduction is smaller than
that obtained by for HYP fermions by mean-field improvement.
As can be seen from Eq.~(\ref{eq:mf_BK}), however, reducing
the $c^\text{Lat}_i$ does not necessarily lead to matching factors
closer to unity. A better measure is the range of the matching
coefficients, i.e. the largest difference between them, for the
continuum and logarithmic contributions then cancel.
This quantity is also given in the Table, and one sees that
it is moderately reduced by improving the gauge action.

\begin{table}[ht] 
\center 
\begin{tabular}{lrrrrrr}
\hline\hline
	& $(a)$  &$(b)$  &$(c)$  &$(d)$ & $(e)$ & $(f)$  \\ 
\hline 
Gluon action & Wilson & Sym & Wilson & Sym & Wilson & Sym \\
Quark action & Naive & Naive & HYP & HYP & HYP & HYP \\
Mean-field impr.? & Y & Y & N & N & Y & Y \\
\hline 
$c^\text{Lat}_{1}   $   & -2.35 & -2.49 & -4.98 &-3.65 &-2.17 &-1.72 \\
$c^\text{Lat}_{2}   $   &-12.92 &-11.53 &-11.17 &-8.58 &-5.49 &-4.73 \\
$c^\text{Lat}_{3}   $   & -2.95 & -3.08 & -5.50 &-4.12 &-2.69 &-2.19 \\
$c^\text{Lat}_{4}   $   & -3.73 & -2.90 &  1.01 & 1.09 & 1.01 & 1.09 \\
\hline
Range                   & 10.57 &  9.04 & 12.18 & 9.67 & 6.50 & 5.82 \\  
\hline\hline
\end{tabular}
\caption[]{Results for $c^\text{Lat}_i$ for various choices of gauge
and fermion action.
We quote only two decimal places for brevity.}
\label{tab:ff_cLat}
\end{table}

To give a sense of the size of the matching coefficients themselves,
we show in table~\ref{tab:mf_BK} results for the ultrafine MILC
lattices ($a\approx0.045\;$fm), 
setting $\mu=1/a$ (``horizontal matching''), 
and $g^2/(4\pi) = 0.2096$ (the value in the $\overline{MS}$ scheme at $\mu=1/a$).
For the actions we use in practice [column (d)] the one-loop corrections
range between $\pm 8\%$.

\begin{table}[ht]
\center
\begin{tabular}{lrrrrrr}
\hline\hline
	& $(a)$  &$(b)$  &$(c)$  &$(d)$ & $(e)$ & $(f)$ \\ 
\hline 
$c_1$   & 0.978 & 0.980  & 1.022 & 1.000 & 0.975 &  0.968 \\
$c_2$   & 1.154 & 1.131  & 1.125 & 1.082 & 1.030 &  1.018 \\
$c_3$   & 0.988 & 0.990  & 1.031 & 1.008 & 0.984 &  0.975 \\
$c_4$   & 1.001 & 0.987  & 0.922 & 0.921 & 0.922 &  0.921 \\
\hline\hline
\end{tabular}
\caption[]{Values of $c_i$ for the MILC ultrafine lattice.
  Notation for actions as in table~\ref{tab:ff_cLat}.}
  \label{tab:mf_BK}
\end{table}

\section{Conclusion}

We have calculated the matching factors for four-fermion operators 
using various fermion and gauge actions. Most useful are our results
for the fermion action and operators
containing HYP-smeared links,
and with the gluon action being Symanzik improved, for these
are needed for our parallel calculation of $B_K$, and
have been used in Ref.~\cite{Bae:2010ki}.
For these choices, the one-loop corrections are of moderate size, with the
range of corrections being $\approx 10 \times \alpha_s/(4\pi)
\approx \alpha_s$, which is the naively expected size.
The impact of improving the gauge action turns out to be numerically small,
much less than the reduction in the size of the corrections caused by
using mean-field improved operators.

\section{Acknowledgments}
The research of W.~Lee is supported by the Creative Research
Initiatives program (3348-20090015) of the NRF grant funded by the
Korean government (MEST).
The work of S.~Sharpe is supported in part by the US DOE grant
no.~DE-FG02-96ER40956.

\end{document}